\documentstyle[11pt]{article}

\begin{document}

\begin{center}
{\LARGE \bf Hot scalar radiation setting bounds on the curvature coupling parameter}
\\ 
\vspace{1cm} 
{\large V. A. De Lorenci$^{a}$, L. G. Gomes$^{b}$ and E. S. Moreira Jr.$^{b,}$}\footnote{E-mail: 
delorenci@unifei.edu.br, lggomes@unifei.edu.br, moreira@unifei.edu.br}   
\\ 
\vspace{0.3cm} 
{$^{a}$\em Instituto de F\'{\i}sica e Qu\'{\i}mica,}  
{\em Universidade Federal de Itajub\'{a},}   \\
{\em Itajub\'a, Minas Gerais 37500-903, Brazil}  \\
\vspace{0.1cm}
{$^{b}$\em Instituto de Matem\'{a}tica e Computa\c{c}\~{a}o,}  
{\em Universidade Federal de Itajub\'{a},}   \\
{\em Itajub\'a, Minas Gerais 37500-903, Brazil}

\vspace{0.3cm}
{\large April, 2013}
\end{center}
\vspace{1cm}

\begin{abstract}
This paper addresses the interplay between vacuum and thermal local averages
for massless scalar radiation near a plane wall of a large cavity where the
Dirichlet boundary condition is assumed to hold. The main result is that stable
thermodynamic equilibrium is possible only if the curvature coupling parameter is
restricted to a certain range. In more than three spacetime dimensions such a
range contains the conformal coupling, but it does not contain the minimal coupling.
Since this same range for possible values of the curvature coupling parameter
also applies to massive scalar radiation, it may be relevant in settings where
arbitrarily coupled scalar fields are present.

\end{abstract}

\section{Introduction}
When physics of blackbody radiation in a cavity is derived in statistical 
mechanics textbooks the energy of the vacuum is usually 
ignored. The argument is that since the vacuum is by definition the state of minimum
energy of a system, and one is interested only in the energy that can be extracted from the 
cavity, then the energy of the vacuum is effectively zero. In this context, 
electromagnetic radiation in thermodynamic equilibrium with the walls of a large cavity at temperature $T$
is addressed either by considering it as an ideal gas of massless bosons or an ensemble
of harmonic oscillators. Any of these tools leads to the energy density $u$ and pressure $P$
(equation of state) given by,
\begin{eqnarray}
&&
u=\frac{\pi^2}{15}\frac{(k_{B}T)^4}{(\hbar c)^3},
\hspace{1cm}
P=\frac{u}{3},
\label{sm}
\end{eqnarray}
where the fundamental constants $\hbar$ and $c$ in Eq.~(\ref{sm})
point out the quantum-relativistic nature of the phenomenon.

In another different but presumably equivalent approach, one
considers an ensemble of infinite cubic cavities containing each
the quantum field --- quantum field at finite temperature. 
When techniques are applied to the electromagnetic field it results the
following ensemble average for the stress-energy-momentum tensor,
\begin{equation}
\left<{\cal T}^\mu{}^\nu\right> ={\rm diag}(u,P,P,P),
\label{semt}
\end{equation} 
with $u$ and $P$ given precisely by Eq.~(\ref{sm}).
If one of the walls of the cavity is brought isothermally  from infinity
and parallel to the opposite fixed wall, $\left<{\cal T}^\mu{}^\nu\right>$ changes radically, 
loosing its isotropic character although remaining uniform \cite{bro69}.
When the distance between these two walls is small enough,
$\left<{\cal T}^\mu{}^\nu\right>$ is such that hardly depends on $T$; in fact it does not
even vanish when $T=0$. At this stage, the two walls are
attracted to each other --- the Casimir effect --- and that is a surprising dynamical manifestation
of the vacuum state in quantum field theory (see reviews \cite{mil01,bor01}).

There are also manifestations of the electromagnetic vacuum when the walls are far apart. When a large cavity has curved walls \cite{bal78,deu79},
the behaviour of electromagnetic radiation in the bulk
is still essentially given by Eqs.~(\ref{semt}) and (\ref{sm}).
Approaching a wall one sees that $\left<{\cal T}^\mu{}^\nu\right>$, in fact, is not uniform.
In particular near the wall the expression for $u$ in Eq.~(\ref{sm})
becomes a subleading contribution in $\left<{\cal T}^0{}^0\right>$.
The corresponding leading contribution is a temperature independent term, which diverges as the curved wall is approached if the perfect 
conductor boundary conditions are taken on it. 
This non integrable divergence is long known in the literature, 
and it is commonly understood as a consequence of overidealizing  a 
real conductor (see also  \cite{ken80,ken82}, for an alternative interpretation).
When a ``cut off procedure'' is considered to modelling a real
conductor, $\left<{\cal T}^\mu{}^\nu\right>$ remains finite as the wall is approached, 
while keeping its Planckian thermal behaviour.

So far only electromagnetic radiation has been addressed.
Considering now massless scalar radiation,
deep in the bulk of a large cavity Eqs.~(\ref{semt}) and (\ref{sm}) hold,
after halving $u$ (consistent with absence of polarization) \cite{ken80}. 
In analogy with electromagnetic radiation in a cavity with curved walls, one will eventually
notice that $\left<{\cal T}^\mu{}^\nu\right>$ is not uniform by moving toward a wall, 
but this time regardless of whether the wall is curved or not.
As has been shown in  \cite{ken80}, the thermal behaviour
of $\left<{\cal T}^0{}^0\right>$ near a reflecting plane wall will be still Planckian, 
but the corresponding ``Stefan constant''  will depend on the kind 
of boundary condition --- Dirichlet or Neumann --- that is assumed to hold on the wall.
Again, (zero temperature \cite{ken80}) non integrable divergences may appear in  $\left<{\cal T}^\mu{}^\nu\right>$ on the wall, and  ``cut off procedures'' to eventually cure them will have to be judiciously chosen to avoid violation of the principle of virtual work \cite{ful10}.

Another new ingredient that thermal scalar radiation brings about, and that has been overlooked in the literature, is
the thermal dependence of $\left<{\cal T}^\mu{}^\nu\right>$ on the curvature coupling parameter $\xi$. 
As is well known, in flat spacetime  $\xi$ 
does not appear in the wave equation; but it does appear in the 
corresponding stress-energy-momentum tensor. The reason is 
that the variation of the action with respect to the metric that leads
to ${\cal T}^\mu{}^\nu$ is made
before solving the Einstein equations which yield the flat geometry.

Thermal behaviour of $\left<{\cal T}^\mu{}^\nu\right>$
for massless scalar fields near  plane walls in four spacetime dimensions has been investigated in  Refs. \cite{ken80} and \cite{tad86}.
The present work extends the investigation by examining   
the thermal dependence of $\left<{\cal T}^\mu{}^\nu\right>$ 
on $\xi$ near a reflecting plane wall of an infinite cavity in flat spacetime with arbitrary number of dimensions. 
The organization of the paper will be as follows.
In Section (\ref{propagator}) 
the Feynman propagator at finite temperature is obtained, and
used according to the ``point splitting procedure'' \cite{dav82}
to derive, in Section \ref{stensor},
the asymptotic behaviours (i.e., at low and high temperatures)
of $\left<{\cal T}^\mu{}^\nu\right>$. 
An analysis of certain local properties of the hot scalar radiation
is presented  in Section \ref{features}. 
The main result in the paper is in Section \ref{bounds}, namely, 
by requiring stable thermodynamic equilibrium it is shown that
only values of $\xi$ restricted to a certain range are acceptable. 
A few final points are addressed in Section \ref{conclusion}
that also contains a summary.

Before proceeding to the next section, a word of caution is in order.
This paper concerns the thermal behaviour of local quantities
such as the energy density and stresses. It differs from papers 
(e.g., Refs. \cite{dow78,amb83})
that investigate
global quantities, 
among them the Helmholtz free energy and the internal energy.
In principle, global quantities can be obtained by integrating the corresponding local quantities; but as mentioned in Ref. \cite{ken80} 
it would be
erroneous to assume that one can infer the thermal behaviour of the latter
looking at the thermal behaviour of the former.

In the rest of the text, unless stated otherwise, $\kappa_{B}=\hbar=c=1$.


\section{The scalar propagator}
\label{propagator}
A plane wall is taken at $x=0$, 
where the massless scalar field $\phi$ vanishes --- Dirichlet's boundary condition.
All the other walls of the $(N-1)$-dimensional cavity are at infinity. 
Radiation in the cavity is assumed to be in thermodynamic equilibrium with the
wall at temperature $T$, so one uses the formalism of
analytically continuing time $t$ to imaginary values, taking $it$
periodic with period $1/T$. Observing these prescriptions the equation  \cite{dav82}
\begin{equation}
 \Box_{{\rm x}}D_{{\cal F}}({\rm x},{\rm x}')=
-\delta^{N}\left({\rm x}-{\rm x}'\right)
\label{fe0}
\end{equation}
is  solved to obtain the scalar propagator. 
The solution is well known for $N=4$ (see  \cite{dav82}, and references therein). 
When $N>3$,
\begin{eqnarray}
D_{{\cal F}}({\rm x},{\rm x}')=
-\frac{i}{4\pi^{N/2}}\Gamma\left(\frac{N-2}{2}\right)
\sum_{n=-\infty}^{\infty}
\left[(-\sigma_{-})^{(2-N)/2}-(-\sigma_{+})^{(2-N)/2}\right],
&&
\label{pro}
\end{eqnarray}
where 
$\sigma_{\pm}:=(t-t'-in/T)^2-(x\pm x')^2-(y-y')^2-(z-z')^2-\cdots$
plus an infinitesimal negative imaginary term. Noting that the
 term corresponding to $n=0$  and involving $\sigma_{-}$
in Eq.~(\ref{pro}) is simply the zero temperature propagator 
in ordinary Minkowski spacetime, direct application of 
$\Box_{{\rm x}}$ to Eq.~(\ref{pro}) yields promptly  Eq.~(\ref{fe0}).
In the following, the Minkowski propagator 
will be dropped in order to implement renormalization.
It is worth mentioning that for $N\leq 3$ familiar divergences arise \cite{ful87}.
As will be seen shortly, such divergences do not bother 
$\left<{\cal T}^\mu{}^\nu\right>$ though when $N\geq 2$.

\section{The stress-energy-momentum tensor}
\label{stensor}
\- The expectation value of the stress-energy-momentum tensor can be
obtained by applying the differential operator \cite{dav82}
$$
{\cal D}^\mu{}^\nu:=(1-2\xi)\partial^\mu\partial^{\nu'}
- 2\xi\partial^\mu\partial^{\nu} 
+(2\xi - 1/2)\eta^\mu{}^\nu
\partial^\lambda\partial_{\lambda'} 
$$
to the renormalized propagator,
$$
\left<{\cal T}^\mu{}^\nu\right> = i \lim_{{\rm x}'\rightarrow {\rm x}}
{\cal D}^\mu{}^\nu D_{{\cal F}}({\rm x},{\rm x}'),
$$
resulting that  $\left<{\cal T}^\mu{}^\nu\right>$ is conserved, traceless
when $\xi=\xi_{N}$ (conformal coupling),
\begin{equation}
\xi_{N}:=\frac{N-2}{4(N-1)},
\label{cc}
\end{equation}
and with diagonal form,
\begin{equation}
\left<{\cal T}^\mu{}^\nu\right> ={\rm diag}(\rho,P_{\perp},P_{\parallel},\cdots,P_{\parallel}).
\label{stress}
\end{equation} 
The only component that is uniform in Eq.~(\ref{stress})
is the pressure perpendicular to the wall,
\begin{equation}
P_{\perp}=\frac{1}{\pi^{N/2}}
\Gamma\left(\frac{N}{2}\right)
\zeta(N)T^{N}.
\label{pep}
\end{equation}
The interest here is on the asymptotic behaviours of the quantities in 
Eq.~(\ref{stress}). For convenience formulas will be specialized to non negative $x$. 
In the bulk ($Tx\gg 1$) one finds for the energy density,
\begin{equation}
\rho=(N-1)P_{\perp}+\rho_{{\tt class}},
\label{ed}
\end{equation}
where
\begin{eqnarray}
\rho_{{\tt class}}:=\frac{2^{2-N}}{\pi^{(N-1)/2}}(N-2)
\Gamma\left( \frac{N-1}{2}\right)
\left(\xi-\frac{1}{4}\right)\frac{T}{x^{N-1}},
&&
\label{ced}
\end{eqnarray}
and for the pressure parallel to the wall ($N\geq 3$),
\begin{equation}
P_{\parallel}=P_{\perp}+P_{\tt class},
\label{pap}
\end{equation}
where $P_{{\tt class}}$ is given  by the negative of  Eq.~(\ref{ced}) with
$\xi-1/4$ replaced by $\xi -\xi_{N-1}$.
Eqs.~(\ref{ed}) and (\ref{pap}) are exact up to exponentially small corrections,
and at high temperatures they apply also near the wall.

Moving now near the wall ($Tx\ll1$), $P_{\perp}$ is still given by Eq.~(\ref{pep})
since it is uniform. Neglecting terms with higher powers of $Tx$, it results that
\begin{eqnarray}
\rho=\frac{2}{(4\pi)^{N/2}}(N-1)
\Gamma\left(\frac{N}{2}\right)
(\xi-\xi_N)x^{-N}
\hspace{2.0cm}
&&
\nonumber
\\
+\frac{1}{\pi^{N/2}}
\Gamma\left(\frac{N}{2}\right)
\zeta(N)(1-4\xi)T^{N},
\hspace{0.5cm}
&&
\label{edw}
\end{eqnarray}
and that
\begin{equation}
P_{\parallel}=-\rho.
\label{es}
\end{equation}
Clearly, Eqs.~(\ref{edw}) and (\ref{es}) hold also everywhere at low temperatures.

Part of the material in Ref. {\cite{ken80}} concerns with thermal scalar radiation
near a reflecting plane wall.
By setting in Eq.~(\ref{ed}) 
$N=4$ and $\xi=0$ (minimal coupling), or $\xi=1/6$ 
[conformal coupling when $N=4$, cf. Eq.~(\ref{cc})], the 
results in  \cite{ken80} are successfully reproduced. 
Equation (\ref{edw}) also reproduces the corresponding expression in \cite{ken80},
when $N=4$ and $\xi=1/6$. 
Reference \cite{tad86} has investigated thermal scalar radiation between two parallel reflecting walls separated by unity, providing a formula for $\left<{\cal T}^\mu{}^\nu\right>$
with arbitrary $\xi$ and $N=4$. 
It has been checked that, by reintroducing arbitrary 
distance between the walls and taking it to infinity, 
the resulting asymptotic behaviours of $\left<{\cal T}^\mu{}^\nu\right>$
in  \cite{tad86} agree with those above, after setting $N=4$.

\section{Some local features of the hot radiation} 
\label{features}

The expression for $P_{\perp}$ in Eq.~(\ref{pep}), which holds everywhere
at arbitrary $T$, is precisely the usual blackbody radiation pressure
in $N$ dimensions. 
Deep in the bulk ($x\rightarrow\infty$) 
$\rho_{{\tt class}}$ and $P_{{\tt class}}$ in Eqs.~(\ref{ed}) and (\ref{pap})
can be neglected, resulting the usual relations proper of the uniform and isotropic
blackbody radiation. 

By reintroducing
dimensionful $\hbar$ in Eqs.~(\ref{ed}) and (\ref{pap}),  one sees that 
the ``classical'' corrections
$\rho_{{\tt class}}$ and $P_{{\tt class}}$ carry $\hbar^{0}$ whereas the 
blackbody quantities carry a negative power of $\hbar$, namely,
$\hbar^{1-N}$. 
Although the terminology
``classical'' is at some extent justified, by setting
$\hbar\rightarrow 0$ the ``classical'' corrections effectively vanish since 
the blackbody quantities diverge. 
In the context of $N=4$,  Ref. \cite{ken80}
has pointed out that the only relevant correction to the blackbody 
contribution in Eq.~(\ref{ed}) is linear in $T$. Now it can be seen from
Eq.~(\ref{ced}) that this is so
regardless the number of dimensions $N$. 
This fact resembles the equipartition principle of energy, 
and it could not be reached simply by using dimensional arguments.
At this point it should be noted that  
``classical'' contributions also appear in the context of Casimir's effect at
finite temperature \cite{bro69,bal78} (see also  \cite{mil01,bor01}, and references therein).

Whereas only the subleading contribution in the expression for $\rho$ 
in the bulk [cf. Eq.~(\ref{ed})] carries dependence on $\xi$, both terms 
in Eq.~(\ref{edw})
depend on it. As the wall is approached, the vacuum energy density  in Eq.~(\ref{edw})
diverges for $\xi\neq\xi_{N}$ \cite{rom02}, 
an well known fact.
Such a limitation can be seen as a consequence
of replacing a real reflecting plane wall (which has some thickness) 
by a boundary condition on a plane \cite{deu79}. 
Roughly speaking, certain ``cut off procedures'' result in inserting a small parameter $\epsilon$
in the formula for the  vacuum energy density in Eq.~(\ref{edw}), such that when 
$x=0$ (keeping $\epsilon\neq 0$) the final formula is finite. It follows also that when
$x\gg\epsilon$
(with $Tx\ll 1$ still holding),  
Eq.~(\ref{edw}) is taken as a good approximation.
Another way of tackling the non integrable divergence 
on the reflecting wall is the ``renormalization procedure''
proposed in Ref. \cite{ken80}. According to this procedure
the zero temperature contribution in Eq.~(\ref{edw}) holds literally
out of the wall, but a $\delta$-function contribution is added to it
giving rise to a surface energy that cancels the troublesome divergent contribution in the total
vacuum energy after integration is performed over space.     
 (This ``renormalization procedure'' has successfully  been extended to 
the electromagnetic field in Ref. \cite{ken82}.)

Unlike the vacuum energy density, the ``Planckian contribution''
in  Eq.~(\ref{edw}) is divergence free, and this fact has been stressed in Ref. \cite{ken80}.
Although such a temperature dependent contribution does not depend on $x$, strictly speaking
it should not be taken as purely thermal. Its mixed nature (vacuum-thermal) is reflected in 
the fact that it depends both on $\xi$ and $T$.
As has been pointed out in  \cite{ken80} for $N=4$, and mentioned earlier 
in the text, 
the behaviour in Eq.~(\ref{edw}) resembles that for 
electromagnetic radiation near a curved wall where the 
perfect conductor boundary condition is assumed to hold \cite{bal78}. 
An important difference though is that the subleading contribution in the
case of the electromagnetic radiation is the very blackbody energy density,
and thus purely thermal. Another feature near the wall (or everywhere at low temperatures)
that should also be remarked is that whereas $P_{\perp}$ in Eq.~(\ref{pep}) is always positive,
each term of $P_{\parallel}$ in Eq.~(\ref{es}) can be negative depending on $\xi$.

The expression for the ``specific heat" per unit of volume 
in the bulk
[cf.  Eq.~(\ref{ed})] is given by (quotation marks stands for the fact that $c_{V}$ below is simply the rate of change with temperature of the energy density, rather than the rate of change with temperature of the total energy in the cavity divided by its volume)
\begin{equation}
\hspace{-0.5cm}   c_{V}=\frac{N(N-1)}{\pi^{N/2}}\Gamma\left(\frac{N}{2}\right)
\zeta(N)T^{N-1}+\frac{\rho_{{\tt class}}}{T},
\label{shb}
\end{equation}
where the factor multiplying $NT^{N-1}$ is the familiar
``Stefan constant'' in $N$ spacetime dimensions.  
Near the wall [cf. Eq.~(\ref{edw})],
\begin{equation}
c_{V}=\frac{N}{\pi^{N/2}}\Gamma\left(\frac{N}{2}\right)
\zeta(N)(1-4\xi)T^{N-1},
\label{shw}
\end{equation}
which also has Planckian form, but now the ``Stefan constant''
depends on $\xi$. Examining these equations one sees that
$c_{V}$ in the bulk (or everywhere at high temperatures) is positive, 
but near the wall (or everywhere at low temperatures) it can be negative.

\section{Bounds on $\xi$}
\label{bounds}

Consider now the local conservation law
of energy and momentum, 
\begin{equation}
{\cal T}^\mu{}^\nu,_{\nu}=0.
\label{emconservation}
\end{equation}
In order to make the discussion that follows regarding the allowed values of $\xi$
consistent with stable thermodynamic equilibrium
as clear as possible, the four dimensional case will be treated first.
Thus, in Eq.~(\ref{emconservation}), $\mu$ and $\nu$ run from 0 to 3 corresponding to 
$t$, $x$, $y$ and $z$, respectively. The energy flux density
${\bf S}=({\cal T}^0{}^1,{\cal T}^0{}^2,{\cal T}^0{}^3)$ and the
momentum ${\bf P} $ in a volume $V$ are related by,
\begin{equation}
{\bf P} =\int_{V}{\bf S}\ d^{3}\hspace{-0.05cm}x.
\label{momentum}
\end{equation}
The global conservation of energy follows from Eq.~(\ref{emconservation}),
\begin{equation}
\frac{d}{dt}\int_{V}{\cal T}^0{}^0 d^{3}\hspace{-0.05cm}x
=-\oint_{S}{\cal T}^0{}^i n^{i} da
=-\oint_{S}{\bf S}\cdot{\bf n}\ da,
\label{econservation}
\end{equation}
where ${\bf n}$ is the outward normal to the boundary $S$ of $V$.
Denoting,
\begin{equation}
{\bf F}=\frac{d{\bf P}}{dt},
\label{force}
\end{equation}
the global conservation of momentum is also obtained from
Eq.~(\ref{emconservation}),
\begin{equation}
{\rm F}^{i}=-\oint_{S}{\cal T}^i{}^j n^{j} da,
\label{mconservation}
\end{equation}
where, as in Eq.~(\ref{econservation}), $i$ and $j$ run from 1 to 3,
and repeated indices indicates summation.

To show that $\xi$ has bounds, one begins
by considering a tiny cubic region of the scalar radiation 
in the cavity, 
and say, facing the walls of the cavity. 
Each side of the cube has area $A$ and is imagined to be contained in a rectangular
parallelepiped with volume $A\delta$, where $\delta$ is as small as one likes. There are, therefore, six of these very thin rectangular 
parallelepipeds. Proceeding, consider the three sides of the cube for which the outward normal ${\bf n}$ coincides with  the
usual unity coordinate vectors ${\bf i}$, 
${\bf j}$ and ${\bf k}$. Below, these sides 
and the corresponding rectangular parallelepipeds  will be labelled 
by subscripts (1), (2) and (3), respectively.

It is further assumed that the temperature $T_{in}$
inside the cube may be slightly different from the temperature 
$T_{out}$ outside, and that these temperatures are low enough for 
Eqs.~(\ref{edw}), (\ref{es}) and  (\ref{shw}) to hold.
Setting $N=4$,
it follows then from 
Eqs.~(\ref{mconservation})  and (\ref{stress}) that the time rates of variation
of the momentum in each rectangular parallelepiped are, 
\begin{eqnarray}
&&{\bf F}_{(1)}=\frac{\pi^2}{90} A
\left(T^{4}_{in}-T^{4}_{out}\right)
{\bf i},
\hspace{0.5cm}
{\bf F}_{(2)}=
-\frac{\pi^2}{90} A
(1-4\xi)
\left(T^{4}_{in}-T^{4}_{out}\right)
{\bf j},
\nonumber
\\
&&
\hspace{2cm}
{\bf F}_{(3)}=
-\frac{\pi^2}{90} A
(1-4\xi)
\left(T^{4}_{in}-T^{4}_{out}\right)
{\bf k}.
\label{parallelepipeds}
\end{eqnarray}
If at some instant ${\bf S}$ vanishes
inside the tiny cubic region, after a short time interval $dt$,
\begin{equation}
{\bf S}_{(i)}=\frac{1}{A\delta}{\bf F}_{(i)} dt,
\label{energyflux}
\end{equation}
where Eqs.~(\ref{momentum}) and (\ref{force}) have been used.
Corresponding to Eqs.~(\ref{parallelepipeds}) and
(\ref{energyflux}), the energy flux per unit
of time out of the cube, $\oint_{S}{\bf S}\cdot{\bf n}\ da$, is 
given by 
\begin{equation}
\Phi= \left[1-2(1-4\xi)\right]\left(T^{4}_{in}-T^{4}_{out}\right),
\label{flux}
\end{equation}
up to an overall positive factor. 

Now, for instance, suppose that
$T_{in}>T_{out}.$
If $\xi>1/4$, Eq.~(\ref{flux}) yields $\Phi>0$, and then
the energy inside the tiny cube decreases [cf. Eq.~(\ref{econservation})].
Noting Eq.~(\ref{shw}), when $\xi>1/4$, 
$c_{V}<0$ resulting that $T_{in}$ becomes even higher and $T_ {out}$ lower, i.e.,
the system is taken away  from thermodynamic equilibrium.
It follows that $\xi$ must not be greater than $1/4$.
As $T_{in}>T_{out}$ and $\xi<1/4$
by assumption, the energy inside the cube must decrease 
[cf. Eq.~(\ref{shw})], otherwise the system will run
away from thermodynamic equilibrium. 
Accordingly,  $\Phi$ in Eq.~(\ref{flux}) must be positive, i.e.,
$1-2(1-4\xi)>0$.
Putting all together, $\xi$ needs to be such that
\begin{equation}
\frac{1}{8}<\xi<\frac{1}{4}.
\label{3int}
\end{equation} 

The whole argument can be carried to arbitrary $N$ in a 
straightforward way. In so doing, Eq.~(\ref{flux}) is replaced by
$\Phi=[1-(N-2)(1-4\xi)](T_{in}^{N}-T_{out}^{N})$. Thus, by requiring 
stable thermodynamic equilibrium, one finds that  
$\xi$ needs to be such that
\begin{equation}
\xi<\frac{1}{4}
\label{1int}
\end{equation}
for $N=2$, and
\begin{equation}
\xi_{N-1}<\xi<\frac{1}{4}
\label{2int}
\end{equation} 
for $N>2$.
It is rather curious that the lower bound in Eq.~(\ref{2int}) is the conformal
coupling in one less dimension [cf. Eq.~(\ref{cc})], and that the upper bound is
the limit of $\xi_{N}$ as $N\rightarrow\infty$.
Thus the conformal coupling $\xi_{N}$ fits in Eq.~(\ref{2int}),
but the minimal coupling $\xi=0$ does not. 
As can be readily seen,
to eventually including equalities
in Eqs.~(\ref{1int}) and (\ref{2int}), further corrections in Eqs.
(\ref{edw}) and (\ref{es}) are required. 
It should be additionally noted
that when $N\rightarrow\infty$, the interval in Eq.~(\ref{2int})
narrows to a point, encapsulating the conformal coupling.

A 
connection 
between Eq.~(\ref{2int}) and the ``classical'' corrections in 
Eqs.~(\ref{ced}) and (\ref{pap}) is worth pointing out.
The ``classical'' corrections  
are related by 
\begin{equation}
\left(\xi-\frac{1}{4}\right)P_{\tt class}=\left(\xi_{N-1}-\xi\right)\rho_{\tt class},
\label{ces}
\end{equation}
and therefore Eq.~(\ref{2int}) implies that $P_{\tt class}$ and $\rho_{\tt class}$ have the same sign.
Now, whereas Eq.~(\ref{ces}) is a feature
of the behaviour of the scalar radiation in the bulk
(or everywhere at high temperatures), the permissible 
range in Eq.~(\ref{2int}) arises by considering its behaviour near the wall
(or everywhere at low temperatures).

Mathematically speaking, a wall is a boundary condition assumed
to hold on a surface to  prevent flux of energy 
across the surface. As Eqs.~(\ref{1int}) and (\ref{2int}) were obtained
assuming the Dirichlet boundary condition, 
it is pertinent to ask if another commonly used boundary condition,
namely the Neumann boundary condition, leads to the same result.
In other words one may wish to know how much dependent on the type of
boundary condition Eqs.~(\ref{1int}) and (\ref{2int}) are.
A preliminary answer can be obtained from  \cite{tad86}.
After manipulating formulas in  \cite{tad86} concerning
the Neumann boundary condition, one finds as permissible range
$-5/4<\xi<7/8$,
which is to be compared with Eq.~(\ref{3int}).
It follows that, in this case, the permissible range
corresponding to the Neumann boundary condition
not only includes the minimal coupling $\xi=0$, but also contains the range
corresponding to the Dirichlet boundary condition. The Dirichlet boundary
condition is therefore more restrictive. 

Another pertinent question that could be raised is
how much dependent on the mass $m$ of the scalar field  
Eqs.~(\ref{1int}) and (\ref{2int}) are. It is expected that
for high temperatures ($T\gg m$) 
massive scalar radiation behaves as if it were
massless, and therefore Eqs.~(\ref{1int}) and (\ref{2int}) 
should hold. In fact, it has been checked that these equations do hold for arbitrary
$m$. When $m\neq 0$, modified Bessel functions
of the second kind come into play, but they do so without affecting the range of 
permissible values of $\xi$. 

\section{Final remarks} 
\label{conclusion}

An important issue to be addressed here is to which extend 
the atmosphere near the wall can be considered as an ideal 
gas of massless scalar bosons  at temperature $T$.
Taking for convenience $N=4$, one may ask to which extend a standard formula
such as (reintroducing dimensionful constants)
\begin{equation}
\rho=
\int \frac{d^{3}p}{h^3} \frac{pc}{e^{pc/k_{B}T}-1}
\label{u}
\end{equation}  
still holds in the present situation [cf. Eq.~(\ref{stress})].
It should be noted that the right hand side of
Eq.~(\ref{u}) is in fact equal to $U/V$, and strictly speaking the equality might hold only 
if the internal energy $U$ were uniformly distributed in the cavity,
which is clearly not the case for the scalar radiation.
Nevertheless, deep in the bulk ($x\rightarrow \infty$) Eq.~(\ref{u}) and its
corresponding equation of state $p=\rho/3$ hold as follows from the discussion
opening Section \ref{features}. Similarly  
one might guess that the temperature dependent term in Eq.~(\ref{edw})
[see also Eq.~(\ref{shw})],
which holds near the wall (or everywhere at low temperatures), 
could be associated to an effective Planck's distribution as in Eq.~(\ref{u}).
However, recalling how pressure of a gas on the walls of a cavity is obtained from
the distribution function,
one sees clearly that such an effective Planck's distribution would fail in
reproducing  the pressure in Eqs.~(\ref{pep}) and (\ref{es}). This conclusion 
has to do with the already mentioned mixed nature (vacuum-thermal) of the scalar radiation 
at finite temperature near the reflecting wall.

Perhaps this duality can be  better appreciated 
by saying, as in  \cite{bal78},  that the atmosphere near the wall is a mixture of
virtual and real bosons.
Indeed, if initially at any point in the cavity the ``specific heat"
is given by the nearly blackbody expression in Eq.~(\ref{shb}), and then the
temperature is lowered such that eventually $c_{V}$ will be given by  Eq.~(\ref{shw}),
as $T\rightarrow 0$,  $c_{V}\rightarrow 0$ and $P_{\perp}\rightarrow 0$, 
resulting that at $T=0$ only virtual bosons are left
in the cavity from where no further energy can be extracted ---
only vacuum energy is left behind.

Differently from the scalar radiation, the thermal behaviour of the electromagnetic 
energy density $\rho$ in the bulk and near a curved wall of a large cavity 
is the same \cite{bal78}.
Thus an expression like (precisely twice that in) 
Eq.~(\ref{u}) yields the thermal contribution in the bulk and also near the wall,
corresponding to the usual radiation pressure [cf. Eq.~(\ref{sm})].
It follows then that, in this case, 
one can think of an ideal gas of massless bosons (photon gas) 
in the bulk, as well as near the wall. 

It should be noticed that
by dropping the vacuum energy density,
one can check that the integral over space
of $\rho$ in Eq.~(\ref{stress}) does not depend  on $\xi$, for $N>2$. 
This has been pointed out
also in  Ref. \cite{tad86} when $N=4$, for a configuration of two parallel walls, and it is
consistent with the fact that standard blackbody expressions do not
carry dependence on $\xi$.
(Clearly, if the vacuum energy density is included in the integrand, a ``cut off procedure'' must be used to deal with the non integrable divergence
\cite{deu79,ken80,ful10}.)

It is worth pointing out that 
to derive  Eqs.~(\ref{1int}) and (\ref{2int})
the expressions in Section \ref{stensor} were assumed to hold in a situation slightly out of thermodynamic equilibrium, since the temperatures 
$T_{in}$ and $T_{out}$ in Section \ref{bounds} were allowed to be slightly different from each other. 
This assumption is entirely plausible and, indeed, it is the kind of assumption one makes to 
show that the familiar electromagnetic radiation is in stable thermodynamic equilibrium in a cavity. 

Recapitulating, this paper investigated locally the va\-cuum-thermal nature
of the scalar radiation. The main conclusion is that 
statistical mechanics of a scalar field restricts
the values of the curvature coupling parameter $\xi$. 
Corresponding to the Dirichlet boundary condition on a plane wall, 
such a restriction is characterized by the inequalities in
Eqs.~(\ref{1int}) and (\ref{2int}). 
Further still unclear role seems to be played by the bounds in 
Eq.~(\ref{2int}). For example, the symmetrical appearance of these bounds
in the ``classical'' corrections to the blackbody expressions
in the bulk asks for more investigation. 

A particular interesting issue to address in the future  is
possible effects of a more realistic wall on inequalities such as those in Eqs.~(\ref{1int}) and (\ref{2int}), and the use of the Robin 
boundary condition \cite{ken80,rom02} (which  interpolates between Dirichlet and Neumann) may be a good start. Before ending,
it should be mentioned that the present paper fits in a class of works
which constrain couplings involving
curvature by invoking consistency conditions in various contexts
(e.g., Refs. \cite{ada06,hof08,bri08}).

\section*{Acknowledgements}
\hspace{0.1cm} 
The authors are grateful to L. H. Ford for useful suggestions.
This work was partially supported by the
Brazilian research agencies CAPES, CNPq and FAPEMIG. 
%



\end{document}